\documentclass[preprint,12pt]{elsarticle}

\usepackage{graphicx}
\usepackage{subfigure}
\usepackage{amssymb}
\usepackage{amsmath}
\usepackage{float}

\pdfoutput=1 

\journal{Journal of Crystal Growth}

\begin{document}

\begin{frontmatter}

\title{General aspects of the vapor growth of semiconductor crystals - a study based on DFT simulations of the NH$_3$/NH$_2$ covered GaN(0001) surface in hydrogen ambient}

\author[label1]{Pawe\l{} Kempisty}
\address[label1]{Institute of High Pressure Physics, Polish Academy of Sciences, Sokolowska 29/37, 01-142 Warsaw, Poland}
\ead{kempes@unipress.waw.pl}

\author[label1]{Pawe\l{} Strak}
\ead{strak@unipress.waw.pl}

\author[label1]{Konrad Sakowski}
\ead{konrad@unipress.waw.pl}

\author[label1,label2]{Stanis\l{}aw Krukowski \corref{sk}}
\address[label2]{Interdisciplinary Centre for Mathematical and Computational Modelling, University of Warsaw, Pawinskiego 5a, 02-106 Warsaw, Poland}
\cortext[sk]{Corresponding author}
\ead{stach@unipress.waw.pl}

\begin{abstract}
Vapor growth of semiconductors is analyzed using recently obtained dependence of the adsorption energy on the electron charge transfer between the surface adsorbed species and the bulk [Krukowski~et~al. J. Appl. Phys. 114 (2013) 063507, Kempisty~et~al. ArXiv 1307.5778 (2013)]. \textit{Ab initio} calculations were performed to study the physical properties of GaN(0001) surface in ammonia-rich conditions, i.e. covered by mixture of NH$_3$ molecules and NH$_2$ radicals. The Fermi level is pinned at valence band maximum (VBM) and conduction band minimum (CBM) for full coverage by NH$_3$ molecules and NH$_2$ radicals, respectively.  For the crossover content of ammonia of about 25\% monolayer (ML), the Fermi level is unpinned. It was shown that hydrogen adsorption energy depends on the doping in the bulk for the unpinned Fermi level, i.e. for this coverage. Surface structure thermodynamic and mechanical stability criteria are defined and compared. Mechanical stability of the coverage of such surfaces was checked by determination of the desorption energy of hydrogen molecules. Thermodynamic stability analysis indicates that initally equilibrium hydrogen vapor partial pressure steeply increases with NH$_3$ content to attain the crossover NH$_3$/NH$_2$ coverage, i.e. the unpinned Fermi level condition. For such condition the entire range of experimentally accessible pressures belongs showing that vapor growth of semiconductor crystals occurs predominantly for unpinned Fermi level at the surface, i.e. for flat bands. Accordingly, adsorption energy of most species depends on the doping in the bulk that is basis of the possible molecular scenario explaining dependence of the growth and the doping of semiconductor crystals on the doping in the bulk.

\end{abstract}

\begin{keyword}


% Journal of Crystal Growth
A1. Computer simulation \sep A1. Surface processes \sep A3. Metalorganic vapor phase epitaxy \sep B1. Nitrides \sep B2. Semiconducting III-V materials

\end{keyword}

\end{frontmatter}

% \linenumbers

\section{Introduction}
\label{sec:intro}

Gallium nitride growth by metal organic chemical vapor deposition (MOCVD) and hydride vapor phase epitaxy (HVPE) is carried out using ammonia (NH$_3$) as source of active nitrogen. Despite implementation of these methods on a massive scale, some aspects of its still remain unclear and detailed understanding of these processes needs additional investigations.
Therefore thorough description of the atomic scale composition, the structure and the processes at GaN(0001) surface is essential for understanding, controlling and manipulation of the ammonia based growth of the nitride crystals and layers.

Previously, first principles calculations, used to obtain GaN surface phase diagrams showed that the thermodynamically stable structures obey $2\times2$ symmetry and are compliant with the \textit{electron counting rule} (ECR) \cite{walle-prl-88,walle-jvs-20,walle-jcg-248,ito-apl-254}. In simple terms, this rule is satisfied if all electrons can be accommodated in surface bonds in such a way that the valence band (VB) and conduction band (CB) states are filled and empty, respectively \cite{pashley-prb-40}. Recently, Walkosz \textit{et al.} \cite{walkosz-prb-85} showed that InN(0001) surface under NH$_3$ exposure does not obey ECR. 

%%As we demonstrate below, this principle can be also broken for GaN(0001) surface under some coverage which is stable although the ECR is not satisfied for this surface.

In the present work we will apply recently obtained results showing that the adsorption energy depends on the electronic charge transfer during process at the semiconductor surface \cite{krukowski-jap-114,kempisty-arxiv}. It was shown that the difference of the Fermi energy between the semiconductor and the surface adsorbed species leads to the charge transfer if the free states are available \cite{krukowski-jap-114}. The energy contribution depends on the states involved in the electron transfer, effectively these are the states close to the Fermi energy and the surface states emerging due to adsorption. In the case when the Fermi level is not pinned at the surface, the adsorption energy depends on the Fermi level in the bulk i.e. doping \cite{krukowski-jap-114} while in the case of the Fermi level pinned, the energy is different for the different pinning states, i.e. the coverage \cite{kempisty-arxiv}. The effect could change the adsorption energy by several electronvolts, as shown for ammonia adsorption at partially hydrogen covered GaN(0001) surface. The present study  discuss the processes involving hydrogen at NH$_3$/NH$_2$ covered GaN(0001) surface is inadvertently at variance with the earlier studies where such dependence was neither identified nor modeled. The incompatibility of the present and the earlier results is additionally related to the use of an insufficient size of the simulated systems previously, usually based on $2\times2$ supercells, which were not able to reproduce some configurations of the surface. Therefore, in our study we focus on systems larger than those previously published. We report the results for the most stable GaN(0001) surface configurations in ammonia-rich conditions i.e. covered with NH$_3$ and NH$_2$ mixture, obtained employing $4\times4$ supercells. In the analysis of these states the thermodynamic and mechanical stability criteria are defined and compared. The thermodynamic approach was then used to determine the NH$_2$/NH$_3$ coverage in function of partial pressure of molecular hydrogen. The consequences of the present result to general understanding of the growth and doping of semiconductor crystals are discussed in detail. 

\section{Computational details}
\label{sec:details}

All calculations described in this paper were performed using density functional theory (DFT) within GGA approximation and Wu-Cohen exchange-correlation potential \cite{wc-prb-73} employing SIESTA code \cite{ordejon-prb-96,soler-jpcm-02}, employing numerical atomic orbitals and pseudo{\-}potential method. The ions were described by Troullier-Martins pseudopo{\-}tentials and nonlinear core corrections (NLCC) were included for the Ga species which had $3d$ electrons explicitly included as valence states. The basic functions were specifically optimized with the simplex method and have the following implementation: N$^{bulk}$-- DZ, Ga$^{bulk}$-- DZP, N$^{surf}$-- TZP, Ga$^{surf}$-- TZP (3d -- DZ), H -- QZP.

\begin{figure}[t]
  \centering
  \includegraphics[width=4.3 cm]{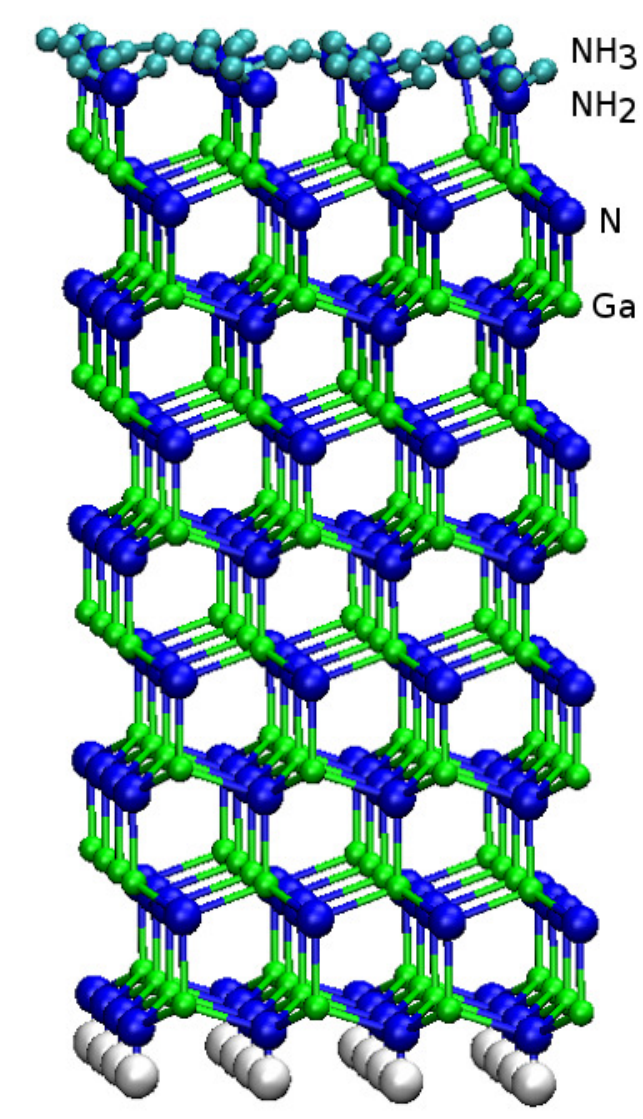}
  \caption{The $4\times4$ slab representing the GaN(0001) surface covered with mixture of NH$_3$ and NH$_2$ admolecules.}
  \label{fig:slab}
\end{figure}

In order to represent the semi-infinite GaN(0001) surface a supercell approach was used with periodically repeated slabs consisting of eight double layers of GaN with lateral size of $4\times4$ units cell, such as the geometry shown in Fig.~\ref{fig:slab}. The vacuum regions of about 47~\r{A} were used to isolate slab replicas. The lower artificial surface of the slab is passivated by pseudo-hydrogen atoms of the fractional atomic number equal to either 0.75 or 0.735. The impact of changes of the slab termination is described in detail in Ref.~\cite{krukowski2-jap-114}. The three topmost layers of atoms are fully relaxed and the all others are relaxed in $c$-direction and fixed at the ideal positions of the lattice $a$. The optimization of geometry was performed until the forces acting on the mobile atoms are smaller than 0.04 eV/\r{A}. Due to the large size of a simulated system ($4\times4$ slab containing more than 320 atoms) only one $k$-point sampling in $\Gamma$ point of first BZ was used during the relaxation of the atomic positions. Then for the final optimized configurations, the total energies were calculated using $3\times3\times1$ Monkhorst-Pack kgrid. The equivalent plane wave cutoff for the grid was set to 275 Ry. The calculated lattice parameters of GaN are $a=3.20$~\r{A} and $c=5.21$~\r{A} in decent agreement with commonly known experimental values: $a = 3.189$~\r{A} and $c = 5.185$~\r{A} \cite{leszcz-apl-69}.

\section{Results}
\label{sec:results}

\subsection{Mechanical and thermodynamic stability}
\label{sec:mt}

In principle the energy calculations determine mechanical stability of the continuum of the possible microstates. Effectively, any mechanical stability criterion may be applied to the energy extrema only. As it is expected, the energy maxima disintegrate instantaneously, thus the energy minima are the only physically relevant states. Mechanical stability is an important class of the stability criteria as it determines whether the considered state (extremum) may survive for a long time. The mechanical stability is usually determined by the energy landscape connecting the two or more energy minima. The one of the lowest energy is the mechanically stable thus it may persist for indefinitely long time, at least in low thermal agitation. 

The higher energy state (extremum) may be metastable in the case of existing energy barrier (then it is a minimum) or unstable in the case when the energy barrier does not exists and the state is attained in the asymptotic sense (effectively it is a neutral energy point). The latter case may by encountered in the desorption processes. Thus the metastable state may survive for a finite time whereas the unstable one disintegrates instantaneously. Therefore the metastable states may be prepared in some, sometimes non-equilibrium conditions to be accessible for investigations. The existence period of such state is determined by the heights of the energy barriers and the thermal agitation of the system. 

Naturally, the mechanically metastable state may last forever under appropriate thermodynamic stimulus compensating the escape from the meta{\-}stable state by the reverse, thermodynamically enforced process. Without this, under thermal agitation, the metastable or even fully stable state may be disintegrated. In the case of the desorption to infinite extent of the vapor, the process may lead to total escape from the surface. Therefore, thermodynamic stability criteria, accounting presence of the compensating flux of the externally chemical potential level, are usually applied to the mechanically stable or metastable states. Depending on these thermodynamic conditions, such surface states may be stabilized, attaining the equilibrium with the extent of infinite vapor or not. 

In summary, mechanical stability criterion determines the possibility of the existence of some surface states. The thermodynamic stability determines which states are achieved in equilibrium with the given chemical potential mass source. In the case of the surface states a physical realization of such sources requires specification of the temperature and the partial pressure of the surrounding vapor. Such procedure is presented below. 

\subsection{Electronic properties}
\label{sec:ep}

\begin{figure*}[t]
  \centering
  \includegraphics[width=12.8 cm]{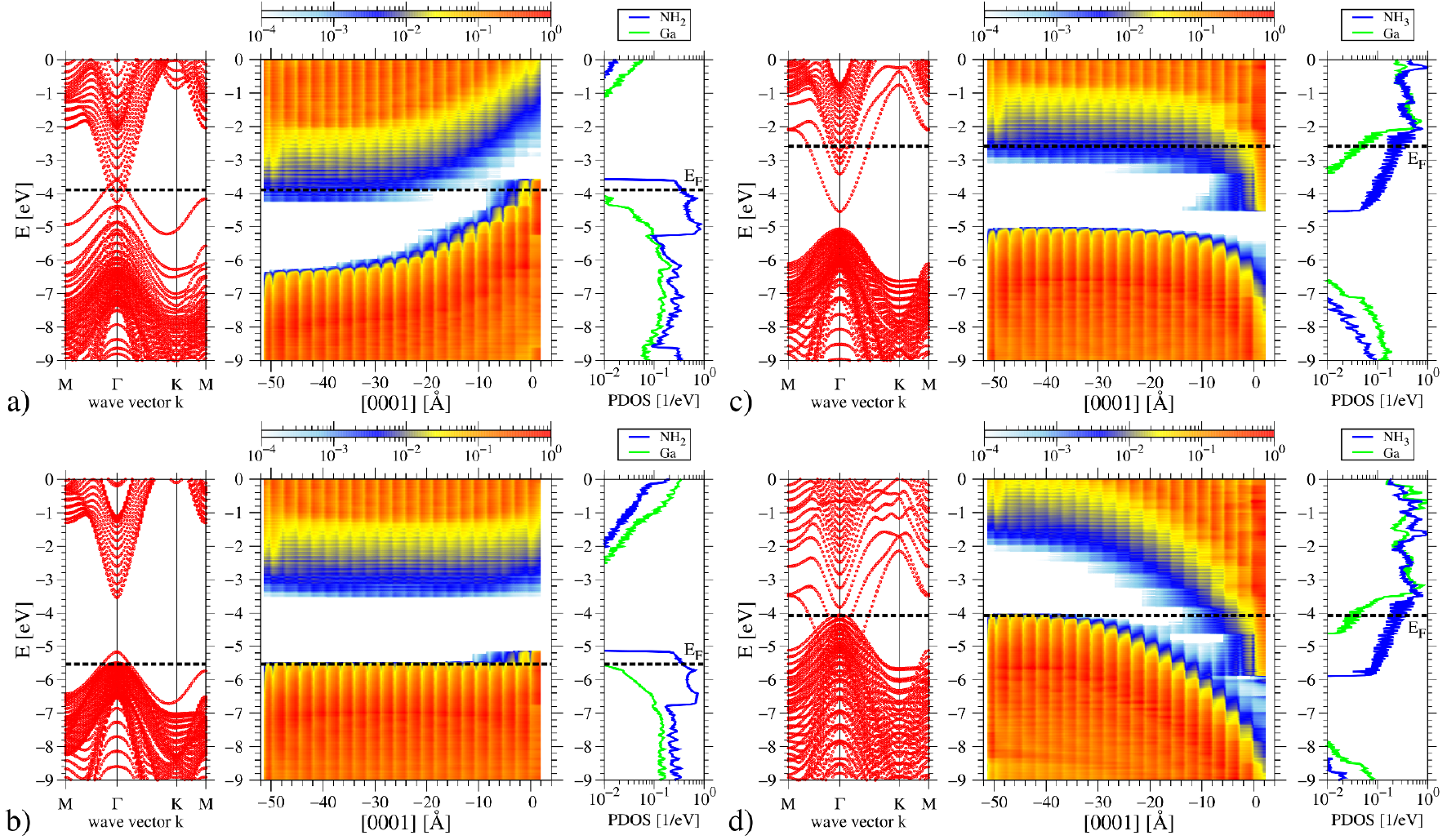}
  \caption{Electronic properties of the ($1\times1$) 20 GaN double atomic layers (DALs) slab representing electronic properties of uniformly NH$_2$ (left column) and NH$_3$ covered (right column) GaN(0001) surface. The diagrams present: left -- the band diagram of the slab, middle -- the density of states (DOS) projected on the atom quantum states (PDOS), showing the spatial variation of the valence and conduction bands in the slab, right -- DOS of the topmost Ga atom (green solid line) and of the N (blue dotted line) and H (grey dashed line) atoms of the adsorbed species. The states density scale (logarythmic) is presented above the diagrams. Top and bottom panels represent $p$- and $n$-type material respectively, that are simulated by charge background procedure implemented in SIESTA package.}
  \label{fig:NH2-NH3}
\end{figure*}
 
In the present investigations, GaN(0001) surface with all sites covered either by NH$_2$ radicals or by NH$_3$ admolecules, is examined. Therefore the surface content of ammonia is defined as number of NH$_3$ molecules divided by total number of the nodes on the slab surface. Due to high partial pressure of ammonia and high V/III ratio, typically close to 4000 in MOVPE growth of GaN layers, it is likely that all Ga sites are covered either by ammonia admolecules or by NH$_2$ radicals. In accordance to earlier publications \cite{walle-prl-88,walle-jvs-20,walle-jcg-248,ito-apl-254} in such dense coverage these molecules are located in the \textit{on-top} positions, as shown in Fig.~\ref{fig:slab}. In the $4\times4$ computational supercell, the maximal number of the surface sites is 16, that are potentially covered by the equal summaric number of ammonia admolecules or NH$_2$ radicals, thus the ammonia coverage can be sampled by 1/16 ML intervals in the range from~0 to 100\%. 
Extreme values mean that GaN(0001) surface are covered uniformly either by NH$_2$ radicals or by NH$_3$ admolecules, with their properties presented in Fig.~\ref{fig:NH2-NH3}. The diagrams show that the Fermi level is pinned at valence band maximum (VBM) and at conduction band minimum CBM) for NH$_2$ and NH$_3$ coverage respectively. Thus, in the case of NH$_2$ coverage, the bands are strongly bend at the surface of $n$-type material (Fig.~\ref{fig:NH2-NH3}a). Naturally the surface state attains considerably high amount of the electronic charge, i.e. becomes surface acceptor. In the case of $p$-type they are almost flat thus the surface is electrically neutral \cite{krukowski2-jap-114}. The second diagram shows the opposite behavior, due to pinning of Fermi level at CBM (Fig.~\ref{fig:NH2-NH3}c,d). The $n$-type bands are close to flat while $p$-type diagram shows large downward bending by positively charged donor state \cite{krukowski2-jap-114}. The presence of such strong bonding suggests that these states are realized experimentally only in extremal external conditions.

\begin{figure}[h]
  \centering
  \includegraphics[width=8 cm]{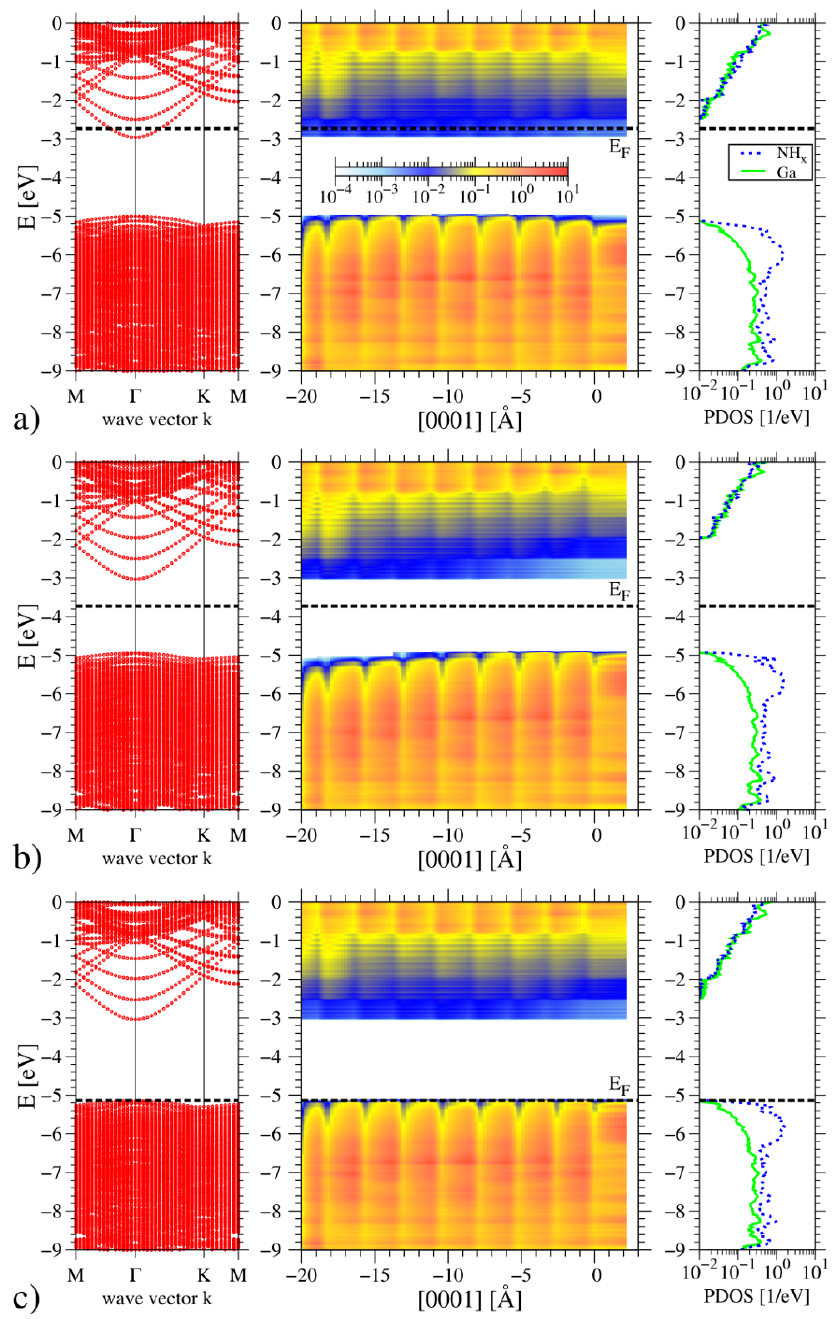}
  \caption{Electronic properties of the ($4\times4$) 8 GaN DALs slab representing electronic properties of 0.25 NH$_3$/0.75 NH$_2$ covered GaN(0001) surface. Left -- the band diagram of the slab, middle -- DOS projected on the atom quantum states (PDOS), showing the spatial variation of the valence and conduction bands in the slab, right -- DOS of the topmost Ga, N and H atoms. The states density scale (logarithmic) is included in the diagrams. Top, middle and bottom panels represent $n$-type, semiinsulating and $p$-type material respectively, simulated by charge background procedure implemented in SIESTA package.}
  \label{fig:bands-nsp-025}
\end{figure}

In the subsequent simulations the ammonia content was consecutively increased by addition of single hydrogen atom, one in every step, to the NH$_2$ radicals. For each such case the relaxation procedure was performed. For all configurations the procedure preserves \textit{on-top} position of the admolecules and radicals. This set of structures was used for investigations of the mechanical stability of the surface, including adsorption and desorption of hydrogen molecules. Naturally, these two above uniform coverage are characterized by differently pinned Fermi levels, thus at some coverage the Fermi level is free at the surface. Such state is attained for 0.25 ML NH$_3$ coverage as demonstrated in (Fig.~\ref{fig:bands-nsp-025}). As it is shown the Fermi level may be shifted across the whole bandgap while the bands remain flat.

\subsection{Hydrogen adsorption and desorption}
\label{sec:h2}

\begin{figure}[h]
  \centering
  \includegraphics[width=8 cm]{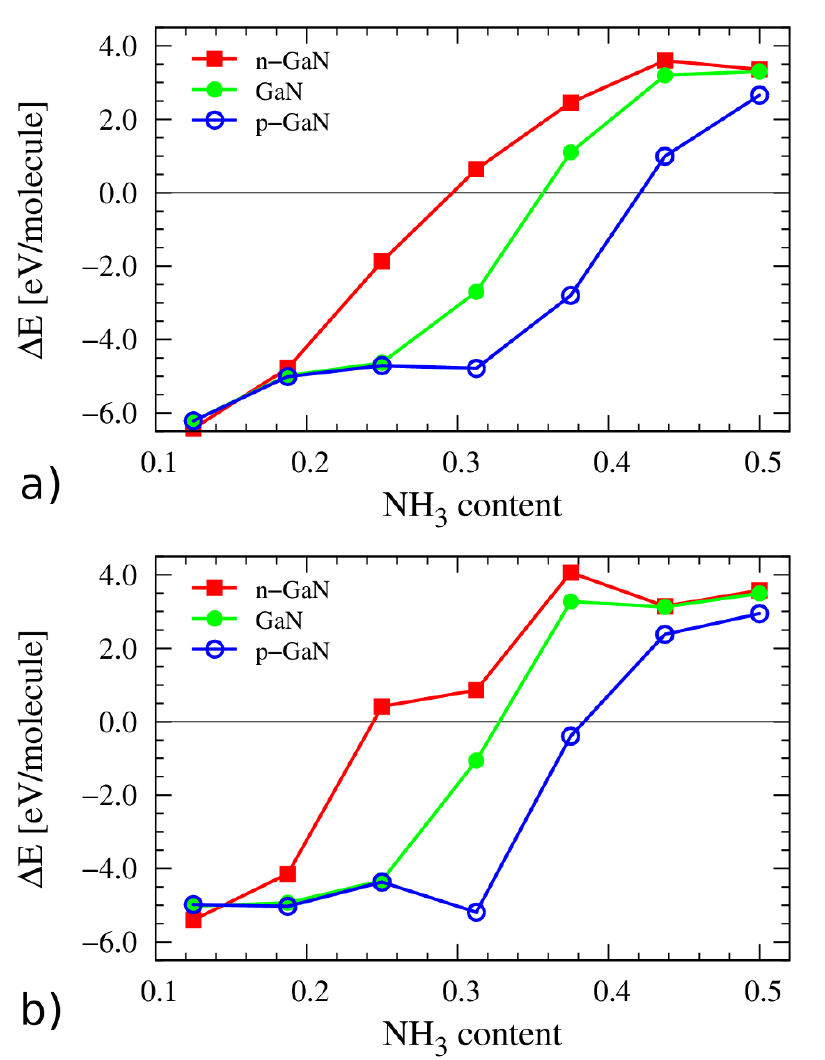}
  \caption{Adsorption energies of the H$_2$ molecule on GaN(0001) surface covered with various mixture of NH$_3$/NH$_2$ obtained from Eq.~\ref{eq:h2} (a) and Eq.~\ref{eq:h1} (b) for $n$-, $p$-type and semiinsulating bulk.}
  \label{fig:h2-3x3x1}
\end{figure}

The hydrogen desorption proceeds by arrangement of the two hydrogen atoms from the neighboring NH$_3$ admolecules that create a bond, i.e. the H$_2$ admolecule which desorbs from the surface. Since, as described above, the relaxed surface structures were already obtained for the configurations having two NH$_3$ admolecules and NH$_2$ radicals respectively, the desorption energy is determined directly as energy difference of these two states:
\begin{equation}
\label{eq:h2}
\Delta E_{H_2} = E_{slab}^{NH_3(i)} - E_{slab}^{NH_3(i-2)} - E_{H_2}
\end{equation}
where the DFT energy of the separate hydrogen molecule $E_{H_2}$ was accounted for. Alternatively, hydrogen removal may be calculated using the data from the desorption processes of single hydrogen atom:
\begin{equation}
\label{eq:h1}
\Delta E_{2H} = 2 \times \left[ E_{slab}^{NH_3(i)} - E_{slab}^{NH_3(i-1)} - E_{H} \right] + E_{diss}^{H_2}
\end{equation}
calculated twice as they create H$_2$ finally so its dissociation energy $E_{diss}^{H_2}$ has to be accounted for. The value taken in the analysis is the hydrogen molecule bonding energy from our DFT calculations $E_{diss}^{H_2}$ = 4.54~eV, slightly different from the experimental value $E_{diss}^{H_2}$ = 4.56~eV \cite{chase}. The index ($i$) denotes the number of NH$_3$ admolecules on the surface. Note that the surface NH$_2$/NH$_3$ coverage is complete, the only considered change here is the relative amount of NH$_3$ admolecules and NH$_2$ radicals. The energies obtained in this way were plotted in Fig.~\ref{fig:h2-3x3x1}.

The difference between the two sets of values obtained from Eq.~\ref{eq:h2} and Eq.~\ref{eq:h1} arises from a finite change of the slab coverages in these two procedures. In the first (molecular) case the change is by 1/8 ML coverage while in the second (atomic) is by 1/16 ML with the energy change doubled. Thus the change in energy is different because the configurations of the slab are different.

\begin{figure*}[t]
  \centering
  \includegraphics[width=12.8 cm]{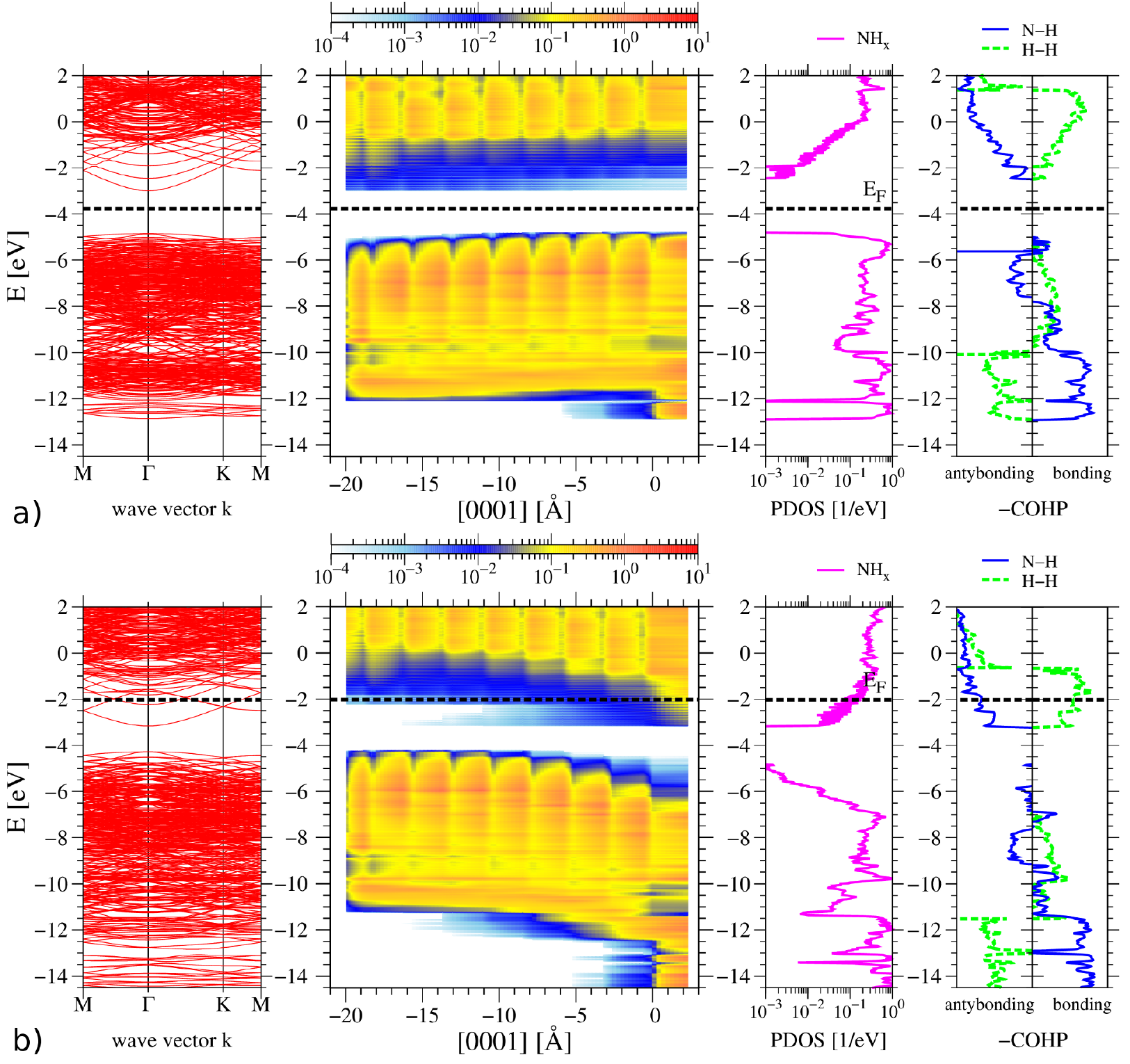}
  \caption{Electronic properties of the ($4\times4$) 8 GaN DALs slab representing electronic properties of 0.25 (a) and of 0.50 (b) NH$_3$ covered GaN(0001) surface; the remaining sites are covered by NH$_2$ radicals. The diagram presents, from left: the band diagram of the slab; DOS projected on the atom quantum states (PDOS); cumulative DOS of NH$_3$ and NH$_2$ species; Crystal Crbital Hamilton Population (COHP \cite{dronskowski-jpc-97}) of H atoms in NH$_3$ molecule and the H and N neighboring atoms. The states density scale (logarithmic) is included in the PDOS diagrams.}
  \label{fig:cohp}
\end{figure*}

The data presented in Fig.~\ref{fig:h2-3x3x1} serve as an example of mechanical stability analysis of the GaN(0001) surface coverage with respect to well defined process of hydrogen desorption connecting two different energy minima. The negative value of $\Delta E$ means that the configuration with the pair of hydrogen atoms attached to the surface have lower energy than that of the H$_2$ molecule separated from the slab. Therefore such a coverage is mechanically stable with respect to molecular hydrogen desorption. The critical, crossover point to unstable region is located in 30\% -- 40\% ammonia content interval depending also on the doping in the bulk. It is worth to underline that the adsorption energy of hydrogen shown in Fig.~\ref{fig:h2-3x3x1} varies from the energy gain of about 5 eV/molecule to the energy loss of about 3 eV/molecule. Thus the total variation of the energy is about 8 eV/molecule i.e. about 4 eV/atom. The change of the adsorption energy can be divided into two components: electron transfer contribution and interaction between the additional hydrogen and the neighboring H and N atoms.
This is verified by the data shown in Fig.~\ref{fig:cohp} where the two configurations differ by 4 hydrogen adatoms, having 4 and 8 NH$_3$ admolecules for top and bottom diagrams respectively. As it is shown, the addition of four H adatoms changes the surface from electrically neutral to acceptor type, confirming that the electronic charge is transferred from the surface states to the bulk. The charge transfer is considerable leading to partial occupation of the conduction band by screening electrons. 

\begin{figure}[h]
  \centering
  \includegraphics[width=8 cm]{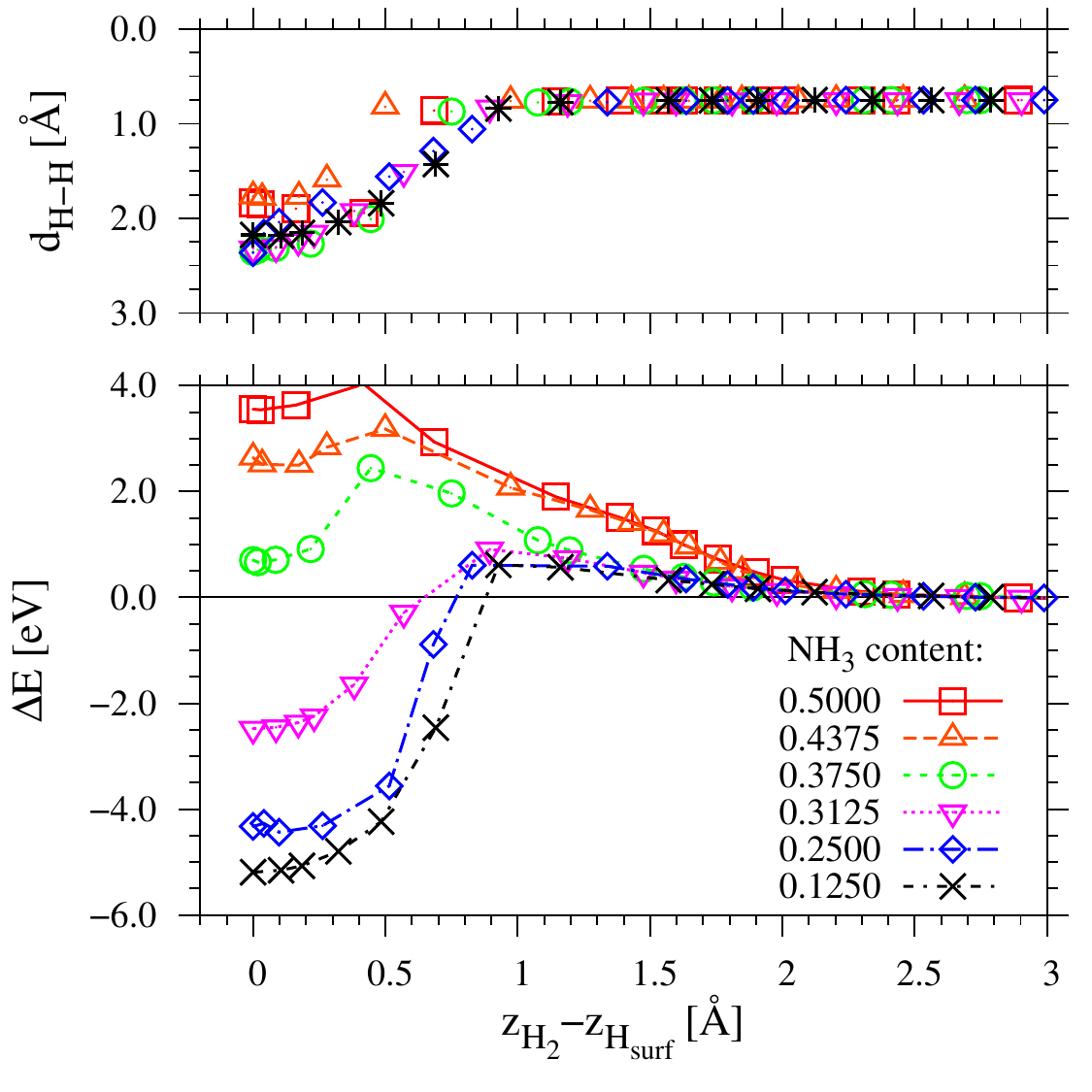}
  \caption{Adsorption/desorption of H$_2$ molecule at GaN(0001) surface under different NH$_3$ coverages obtained by use of NEB method. Top -- interatomic H-H distance, bottom -- the change of the system energy.}
  \label{fig:h2-neb}
\end{figure}

As discussed above, the problem of considerable interest is the metastability of the NH$_2$/NH$_3$ coverage. The magnitude of the energy barriers determines the possible duration of such surface states. Such issue may be solved by a direct determination of the entire path-energy dependence during adsorption (desorption) processes that could verify existence of the energy barrier. The adsorption process was modeled in Born-Oppenheimer approximation by the stepwise motion of H$_2$ molecule from the vacuum far from the slab to the final configuration at the slab surface. The adsorption energy is defined as the difference of the energy of the system at the initial and the final states. For accurate determination of the minimum energy path (MEP) \textit{Nudged Elastic Band} method (NEB) was used \cite{neb-book}.
As expected, during both adsorption and desorption of H$_2$ molecules the energy the barriers are encountered (Fig.~\ref{fig:h2-neb}), obtained as the difference between the transition state and initial energies.

\subsection{Chemical composition of the coverage - the molecular hydrogen pressure dependence}
\label{sec:nh3}

Growth of GaN and InGaN layers by MOVPE is  affected to some extent by the use of hydrogen or nitrogen carrier gas. It was discovered that mere presence of few atomic percent of hydrogen in the vapor decreases indium content in MOVPE grown GaInN layers by one order of magnitude~\cite{bercik-12}. The effect is striking as it is well known that ammonia undergoes thermal decomposition at high temperatures. Therefore GaInN layers are grown using nitrogen gas only, and at relatively low temperatures, not exceeding 1100~K, such that ammonia decomposition is negligible. It is expected that at the typical MOVPE growth arrangement, the decomposition of ammonia in the growth zone is low, so the vapor consist of nitrogen and ammonia with fairly low addition of molecular hydrogen. It is anticipated that the presence of molecular hydrogen may affect the  incorporation of indium into the growing layers either by the change of the coverage of GaN(0001) surface or by the influence on the steps kinetics. The first scenario may be assessed using the above described results.

The principal factor is the thermodynamic stimulus for the adsorption of hydrogen, direct generalization of the supersaturation in standard crystal growth theory, i.e. thermodynamic potential difference between the hydrogen adsorbed at the surface and the chemical potential value of the atomic hydrogen in the vapor: 
\begin{equation}
\label{super_satH}
\Delta \mu_H = \mu[GaN(0001):H] - \mu[GaN(0001) + H]
\end{equation}
or alternatively the pair of hydrogen atoms adsorbed at the surface and the hydrogen molecule $H_2$ in the vapor:
\begin{equation}
\label{super_satH2}
\Delta \mu_{H_2} = \mu[GaN(0001):2H] - \mu[GaN(0001) + H_2]
\end{equation}
The latter describes the physically relevant case of the molecular hydrogen vapor. 

These potential values could be calculated using an extended interpretation of the standard thermodynamic relation between chemical potential, the enthalpy and the temperature/entropy: $\mu = h - Ts$. In the present treatment the $T = 0$~K enthalpy is obtained using calculated above DFT energies. Thus the DFT energy of the slab having ($i$) ammonia admolecules $E_{slab}^{NH_3(i)}$ is used and the chemical potential is then obtained accounting temperature dependent vibrational enthalpy and configurational entropy contributions:
\begin{equation}
\label{en_swob_ads}
\mu[GaN(0001):H] = E_{slab}^{NH_3(i)} + 3 k_B T + k_B T \ln (c)
\end{equation}
The second term describes the vibrational contribution to the enthalpy in the high temperature regime, obtained from equi{\-}partition rule, amounting to $3 k_B T$, and the third one is a configurational entropy term, calculated for $c$ fraction of the ammonia admolecules, equal to $k_B T \ln (c)$. Altogether these two temperature dependent terms are relatively small, at the benchmark GaInN MOVPE temperature $T = 1100$~K, they contribute not more than 0.3~eV, thus their influence is not large.   

Similarly, the chemical potential of atomic hydrogen in the vapor may be expressed using $\mu = h - Ts$ relation, in which as the enthalpy $T = 0$~K reference value, a sum of DFT energy of the slab of ($i$ - 1) ammonia admolecules $E_{slab}^{NH_3(i-1)}$ and the DFT energy separate hydrogen atom $E_{H}$, is taken:
\begin{equation}
\label{en_swob_H}
\mu[GaN(0001)+ H] = E_{slab}^{NH_3(i-1)} + E_{H} + \mu_{H}^{th}(p_H,T)
\end{equation}
The latter term is the temperature and pressure dependent thermodynamic part of chemical potential of atomic hydrogen that can be approximated as \cite{gluszko}:
\begin{equation}
\label{pot_therm_H}
\mu_{H}^{th}(p_H,T) = - 17.29 x - 2.15 x \ln (x)   + k_B T \ln (p_H)
\end{equation}
where $x = T/10^4$, the temperature is expressed in kelvins, hydrogen pressure in bars, and the chemical potential in eV/molecule (atom). This chemical potential attains the zero reference value at 0~K. This quantity is usually used in the thermodynamic stability conditions, e.g. in Refs. \cite{walle-prl-88,walle-jvs-20,walle-jcg-248,ito-apl-254,ito-sst-12,akiyama-jcg-11}.

In the second scenario, the potential of molecular hydrogen H$_2$ in the vapor is used, again employing the standard relation $\mu = h - Ts$: 
\begin{equation}
\label{en_swob_H2}
\begin{split}
& \mu[GaN(0001) + H_2] = E_{slab}^{NH_3(i-2)} + E_{H_2} + \delta \mu_{H_2}^{th}(p_{H_2},T)  \\
& = E_{slab}^{NH_3(i-2)} + E_{H_2} + E_{diss}^{H_2} + \mu_{H_2}^{th}(p_{H_2},T)
\end{split}
\end{equation}
where the first two terms constitute the enthalpy and $\delta \mu_{H_2}^{th}(p_{H_2},T)$ is the temperature and pressure dependent excess thermodynamic part of chemical potential of  molecular hydrogen H$_2$. This quantity, attaining zero level at 0~K, can be approximated as \cite{gluszko}:
\begin{equation}
\label{pot_therm_H2}
\begin{split}
& \delta \mu_{H2}^{th}(p_{H_2},T) = \\
& - 19.96 x - 3.67 x^2 - 2.66 x \ln (x) + k_B T \ln (p_{H_2})
\end{split}
\end{equation}
This parameter differs from a standard thermodynamic potential of molecular hydrogen $\mu_{H2}^{th}(p_{H_2},T)$ by the hydrogen molecule dissociation energy:
\begin{equation}
\label{pot_thermod_H2}
\begin{split}
\mu_{H2}^{th}(p_{H_2},T) = \delta \mu_{H2}^{th}(p_{H_2},T) - E_{dis}^{H_2}
\end{split}
\end{equation}
Naturally, the latter has the reference value at $T = 0~K$, equal to the H$_2$ dissociation energy: $\mu_{H2}^{th}(p_{H_2},0) = - E_{diss}^{H_2}$, and fulfills a standard thermodynamic relation:
\begin{equation}
\label{pot_H_H2}
2 \mu_{H}^{th}(p_{H},T) = \mu_{H2}^{th}(p_{H_2},T)
\end{equation}
Generally, the thermal contribution in the excess thermodynamic potential $\delta \mu_{H_2}^{th}(p_{H_2},T)$  for the vapor is much larger than for the surface and amounts to 1.6~eV at $T=1100$~K. Therefore it affects the equilibrium considerably, shifting the equilibrium towards higher pressures. 

Assumption of the equilibrium between the surface and the molecular hydrogen vapor results in the following condition for chemical potentials $\Delta \mu_{H_2} = 0$ which, accounting Eqs.~\ref{en_swob_H2} and ~\ref{pot_therm_H2}  , gives:
\begin{equation}
\label{equil_press}
\begin{split}
& k_B T \ln (p_{H_2}) = \left[ E_{slab}^{NH_3(i)} - E_{slab}^{NH_3(i-2)} - E_{H_2} \right] \\
&+ 6 k_B T + 2 k_B T \ln (c) + 19.96 x + 3.67 x^2 + 2.66 x \ln (x)
\end{split}
\end{equation}
where the term in the brackets is the DFT molecular hydrogen adsorption energy, given by Eq.~\ref{eq:h2} and plotted in Fig.~\ref{fig:h2-3x3x1}a.
 
Similarly, a parallel derivation of the molecular hydrogen equilibrium pressure may be obtained by use of the atomic hydrogen desorption energy values, given by Eq.~\ref{eq:h1}, and invoking Eqs.~\ref{pot_thermod_H2} and Eq.~\ref{pot_H_H2} :  
\begin{equation}
\label{equil_pres}
\begin{split}
& k_B T \ln (p_{H_2}) = \left[ 2E_{slab}^{NH_3(i)} - 2E_{slab}^{NH_3(i-1)} - 2E_{H} + E_{diss}^{H_2} \right]  \\
&+ 6 k_B T + 2 k_B T \ln (c) + 19.96 x + 3.67 x^2 + 2.66 x \ln (x)
\end{split}
\end{equation}
As before, the term in brackets is the DFT desorption energy of pair of hydrogen atoms corrected by the dissociation energy of the hydrogen molecule, as in Eq.~\ref{eq:h1}, and plotted in Fig.~\ref{fig:h2-3x3x1}b. Note that the pressures derived here are pressure of molecular species, different form the pressures determined in Ref.~\cite{ito-sst-12,akiyama-jcg-11}, where the pressure of atomic vapors were obtained.

\begin{figure}[h]
  \centering
  \includegraphics[width=8 cm]{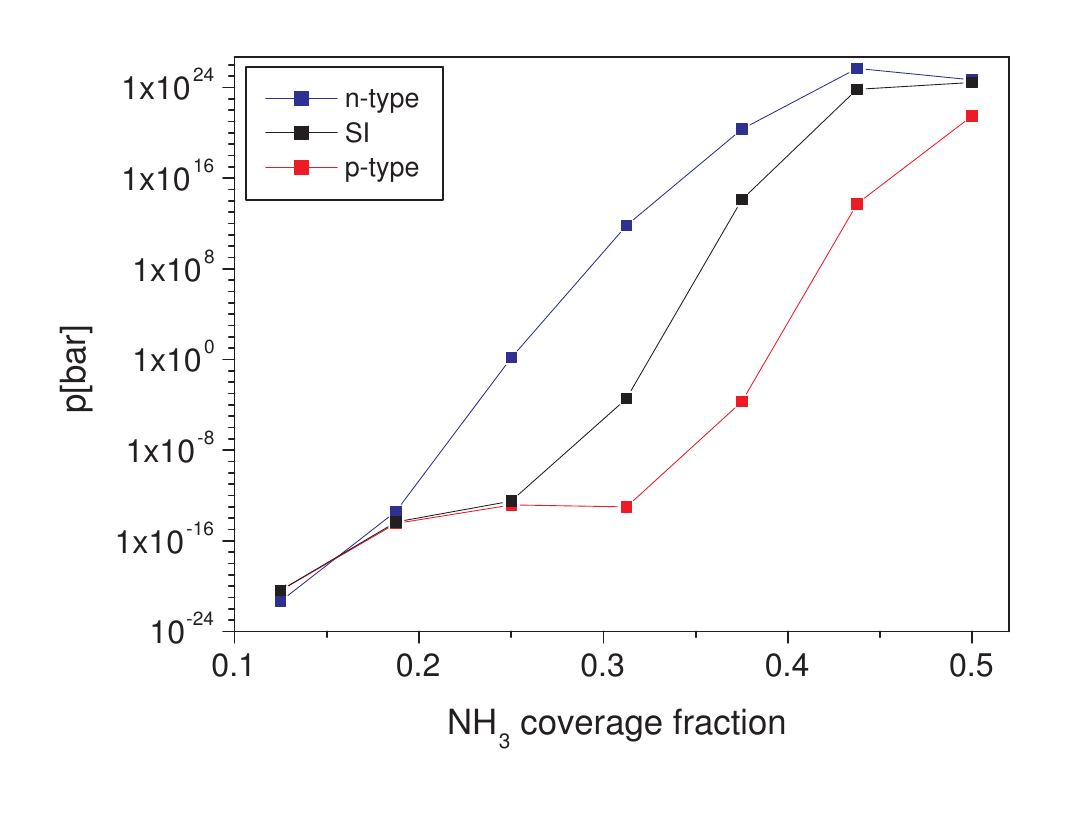}
  \caption{The dependence of mixed NH$_3$/NH$_2$ coverage of GaN(0001) surface on the molecular hydrogen pressure at T = 1100~K. The lines are guiding the eye only.}
  \label{fig:press-H2-npsi}
\end{figure}

These expression were used to determine the relation between the partial pressure of molecular hydrogen and the chemical state of the coverage of GaN(0001) surface. For MOVPE/HVPE growth the vapor phase is dominated by ammonia assuring full coverage of the surface by NH$_3$ molecules and NH$_2$ radicals. The partial pressure of hydrogen affects the relative ratio of these two  components of the coverage in the way presented in Fig.~\ref{fig:press-H2-npsi}. 

It is worth to note that the DFT hydrogen adsorption energies are relatively large and strongly depend on the coverage, changing from -5.2~eV to 3.5~eV, i.e. its total variation exceeds 8~eV. Therefore, the obtained pressures should change by many orders of magnitude for the investigated coverages. 
Accordingly from these data it follows that the pressure necessary to change the coverage in significant way is huge so that the possibility of manipulation of the  NH$_3$/NH$_2$ coverage by control of the partial pressure of molecular hydrogen is negligible. Therefore in the typical MOVPE or HVPE condition the fraction of ammonia molecules is similar and does not exceed 35\%.

Additionally it is worth to note that the hydrogen partial pressure are identical for all types of doping for very low, below 0.2, and very high, above 0.5 NH$_3$ coverage. For the interval between, the partial pressure strongly depends on the doping in the bulk, beeing the larges for $n$-type. Accordingly the respective range of pressure extends from $p_{min} = 10^{-16}$ bar to $p_{max} = 10^{20}$ bar. This interval covers with huge excess the entire pressure range where the growth is possible, assuring that the growth occurs in the condition of the surface with the  Fermi level unpinned, and where the adsorption of the open shell species strongly depends on the Fermi level in the bulk, i.e. the doping. That translates into dependence of the incorporation of the impurities during the growth. In addition to the consequences to GaN growth by ammonia based methods, these results indicate on general features of the growth of semiconductors from the vapor which is formulated in the following Subsection.

\subsection{General features of the growth of semiconductors from the vapor}
\label{sec:vapor}

The above discussed data provide clear guidance for the general scenario of the growth of semiconductors from the vapor. In a staggering majority of cases, the semiconductor's crystal surfaces have Fermi level pinned by surface states \cite{monch-book}. Thus the initial state is Fermi level is pinned by the surface states existing in the absence of the growth species at the surface. The growth proceeds by an increase of supersaturation in the vapor, i.e. the increase of partial pressure of the growth species. They are adsorbed at the surface, changing the energy of the surface states which ultimately should change the energy of the pinning state at the surface. As was shown recently that affects the adsorption energy of the growth species at the surface \cite{krukowski-jap-114,kempisty-arxiv}. On virtue of van t'Hoff relation, the equilibrium pressure increases exponentially with the change of the adsorption energy i.e. effectively the energy of vaporization. Thus the equilibrium pressure changes by 10--20 orders of magnitude as it was shown for hydrogen in Fig.~\ref{fig:press-H2-npsi} so that the initial pinning state is left and the target state is not attained. Therefore in the real vapor growth process the Fermi level is always free. According to the results presented in Ref.~\cite{krukowski-jap-114}, the adsorption energy depends on the Fermi level in the bulk, i.e. on the doping. Thus the novel molecular mechanism, responsible for the dependence of the vapor growth and the doping of semiconductors was identified and described.

These arguments may be summarized as follows:
\begin{itemize}
\item{the adsorption energy of open shell system species depends on the electronic charge transfer at surfaces}
\item{for pinned Fermi level surfaces it depends on the pinning state}
\item{for unpinned Fermi level surfaces it depends on doping in the bulk} 
\item{for majority semiconductor surfaces the Fermi level is pinned}
\item{the growth proceeds by an increase of the density of adsorbed active species, depinning the Fermi level}
\item{the equilibrium pressure of active species is exponentially dependent on the adsorption energy, thus for typical change of the coverage it changes by more than 10 orders of magnitude}
\item{the pressure change during typical growth process is not likely to be sufficient to attain the second pinning state, thus Fermi level is unpinned during the process}
\item{the adsorption energy depends on the Fermi level in the bulk, i.e. the doping that constitutes primary molecular mechanism explaining influence of the doping in the bulk on the vapor growth and the doping of semiconductors} 
\end{itemize}

The above points constitute terse description of the newly proposed scenario explaining basic features of the vapor growth and doping of the semiconductor crystals in general. 

\section{Summary}

The newly proposed dependence of the adsorption energy on the charge transfer at the surface was confirmed. It was shown that the adsorption energy of hydrogen at NH$_3$/NH$_2$ covered GaN(0001) surface changes by several electronvolts depinning on the Fermi level at the surface (pinned) and in the bulk (unpinned). 

The two different stability conditions are analyzed: the newly defined mechanical and the well known thermodynamic. It is argued that the mechanical stability criterion defines unstable regions of the states that are not accessible in the standard experiments. The mechanically stable regions, i.e.~those potentially realized in the experiments are divided into meta- and fully stable parts. The meta and fully stable states may be realized and their existence depends on the thermodynamic state of the surrounding vapor. 

The thermodynamic stability is then invoked and analyzed for the GaN(0001) surface covered by a mixture of NH$_3$ molecules and NH$_2$ radicals with respect to the exchange of hydrogen molecules from the gas phase. It has been shown that the surface becomes mechanically metastable with respect to the escape of H$_2$ molecules at the ammonia content in the 30--40\% interval, depending on the doping in the bulk. The metastable NH$_3$/NH$_2$ covered surface states were shown to depend on the chemical potential of hydrogen in the vapor. By thermodynamic analysis the chemical potential of hydrogen is translated into the thermodynamic control parameters of real processes: the temperature and the partial pressure of molecular hydrogen. It was shown that relatively small change of the content of ammonia admolecules is equivalent to the change of the partial pressures of molecular hydrogen by several orders of magnitude. Thus the impact of hydrogen on the MOVPE GaN growth is limited and hydrogen partial pressure may change the state of the surface only in the limited range of the coverage compositions.

The new general scenario of the growth of semiconductor crystal on the surface is presented in which the Fermi level is unpinned during the process. This observation led to formulation of the novel molecular mechanism explaining the dependence of the growth and doping on the position of the Fermi level in the bulk, i.e. doping in the bulk.

The analysis of the electron density of states shows that GaN(0001) surface under mixed NH$_3$/NH$_2$ coverage does not fulfill electron counting rule for the ammonia content different from 25\%. At these state the Fermi levels penetrates the valence/conduction bands while the surface could be still mechanically stable. Therefore this phenomenon requires further analysis using the methods allowing the investigators to control the electric charge and the electric fields at the surfaces \cite{krukowski2-jap-114}.

\section*{Acknowledgement}
The research was supported by funds of National Science Center of Poland granted by decision No.~DEC-2011/01/N/{\-}ST3/04382.
This research was supported in part by PL-Grid Infrastructure and computing facilities of the Interdisciplinary Centre for Modelling, University of Warsaw (ICM UW), Poland.
Calculations were performed using the SIESTA package \cite{ordejon-prb-96,soler-jpcm-02}. 
Images of atomic configurations were generated using XCRYSDEN shareware \cite{xcrysden}.

\section*{References}

\bibliographystyle{model1a-num-names}
%\bibliography{stability}

\end{document}